# How Sublimation Delays the Onset of Dusty Debris Disk Formation Around White Dwarf Stars

Jordan K. Steckloff[1,2], John Debes[3], Amy Steele[3,4], Brandon Johnson[5,6], Elisabeth R. Adams[1], Seth A. Jacobson[7], Alessondra Springmann[8]

[1]Planetary Science Institute, Tucson, AZ
[2]Dept. of Aerospace Engineering & Eng. Mechanics, University of Texas at Austin, Austin, TX
[3]Space Telescope Institute, Baltimore, MD
[4]Dept. of Astronomy, University of Maryland, College Park, MD
[5]Dept. of Earth, Atmospheric, and Planetary Sciences, Purdue University, West Lafayette, IN
[6]Dept. of Physics and Astronomy, Purdue University, West Lafayette, IN
[7]Dept. of Earth and Environmental Sciences, Michigan State University, East Lansing, MI
[8]Lunar and Planetary Laboratory, University of Arizona, Tucson, AZ

## ABSTRACT

Although numerous white dwarf stars host dusty debris disks, the temperature distribution of these stars differs significantly from the white dwarf population as a whole. Dusty debris disks exist exclusively around white dwarfs cooler than 27,000 K. This is all the more enigmatic given that the formation processes of dusty debris disks should favor younger, hotter white dwarfs, which likely host more dynamically unstable planetary systems. Here we apply a sophisticated material sublimation model to white dwarf systems to show that these statistics are actually a natural result of the interplay of thermal and tidal forces, and show how they define the circumstellar regions where dusty debris disks can form. We demonstrate that these processes tend to prevent stability against both sublimative destruction and reaccretion into planetesimals for rocky materials until white dwarfs cool to below ~25,000–32,000 K, in agreement with the observed limit of ~27,000 K. For pure water ice, this critical temperature is less than 2,700 K (requiring a cooling age older the universe); this precludes pure water ice-rich debris disks forming through the accepted two-step mechanism. The critical temperature is size-dependent; more massive white dwarfs could potentially host dusty debris disks at warmer temperatures.. Our model suggests that the location of the disks within the PG 0010+280, GD 56, GD 362, and PG 1541+651 systems are consistent with a forsterite-dominated olivine composition. We also find that very cool white dwarfs may simultaneously host multiple, independently formed dusty debris disks, consistent with observations of the LSPM J0207+3331 system.

## I. Introduction

A wealth of evidence suggests that white dwarfs, the glowing stellar remnants of dead stars, have actively evolving planetary systems. Between 25-50% of white dwarfs exhibit spectra "polluted" with heavy elements that suggest rocky materials are actively accreting onto the star (*e.g., Zuckerman et al. 2003; 2010; ; Barstow et al. 2014; Koester et al. 2014*). Additionally, ~1-3% of white dwarfs emit detectable excesses of infrared blackbody radiation (*Debes et al. 2011; Rocchetto et al. 2015; Wilson et al. 2019*), which is indicative of a dusty debris disk orbiting within the white dwarf's tidal disruption radius (i.e., Roche Limit; *e.g., Debes & Sigurdsson, 2002; Jura, 2003*). These materials must have migrated inward from greater astrocentric distances (*Debes & Sigurdsson, 2002; Jura, 2003; Reach et al. 2005*) because any sub-Jovian object within a few AU of the star should have been destroyed during the preceding asymptotic giant branch (AGB) stellar phase (*Villaver & Livio, 2007; Nordhaus & Spiegel, 2013*).

This migration has been hypothesized to occur through a two-step process that acts upon planetary objects scattered into highly eccentric, low periapse orbits (*Bonsor et al. 2011; Bonsor & Wyatt, 2012; Debes et al. 2012; Mustill et al. 2018*). First, stellar tidal forces disrupt such objects in orbits with sufficiently small pericenters into dusty debris (*Graham et al. 1990; Jura, 2003; Debes et al. 2012; Veras et al. 2014; Malamud and Perets, 2020* ), which resists radiation pressure from the white dwarf given its low luminosity (*Farihi et al. 2008*). Subsequently, Poynting-Robertson (P-R) drag and other drag processes would rapidly circularize the orbits of the dusty debris, forming a debris disk whose material can spiral inward, vaporize, and pollute the white dwarf (*Hansen et al. 2006; Veras et al. 2015; Farihi, 2016; Malamud et al. 2021*).

However, such dusty debris disks seem to selectively appear only around white dwarfs cooler than ~27,000 K (*e.g., Xu & Jura, 2012; Xu et al. 2015; Li et al. 2017; Debes et al. 2019*), which corresponds to a cooling age of ~17 Myr for a ~0.6 $M_{\odot}$ white dwarf (*e.g., Xu et al. 2015*). This statistic is highly inconsistent with expectations in the two-step mechanism for forming dusty debris disks. Young, hot white dwarfs are expected to inherit the most dynamically active planetary systems (*Debes & Sigurdsson, 2002; Veras et al. 2013; Frewen & Hansen, 2014*), since mass loss at the end of the progenitor star's life could massively destabilize any system of planets or asteroids. Thus, inward scattering and the resulting formation of debris disks should preferentially favor debris disk formation around hot white dwarfs.

Previous works have suggested that the interplay between tidal and sublimative processes create regions where dusty debris is stable against sublimation and reaccretion (*von Hippel et al. 2007; Koester et al. 2014*). Here we introduce a sophisticated material sublimation model (*Steckloff et al. 2015; Steckloff and Jacobson, 2016; Springmann et al. 2019)* to investigate this interplay and show that dusty debris statistics are compatible with this two step formation process, where the restriction of dusty debris disks to cooler white dwarfs (≲27,000 K, cooling ages >~16 Myr) is a natural result of the physical processes controlling dusty debris disk formation.

## II. Physical Processes Constraining Dusty Debris Disks

Physical processes define the circumstellar regions where debris disks can stably reside. The outer edge of debris disks is controlled by stellar tides, which disrupt inwardly scattered planetesimals into dusty debris (*e.g, Jura, 2003; Reach et al. 2005; Debes et al. 2012; Veras et al. 2014; Farihi, 2016; Malamud and Perets, 2020*). Meanwhile, the inner edge of the debris disk is controlled by the thermal environment created by the white dwarf, which can sublimate/vaporize disk materials if extreme enough (*von Hippel et al. 2007*). Thus, the region where dusty debris disks can exist is *outside* of this sublimation radius, yet *inside* of the Roche limit. To compute the astrocentric distances of these two limits, we use standard formulations of Roche limits and a model of material sublimation based on first principles (*Steckloff et al. 2015; Steckloff & Jacobson 2016; Springmann et al. 2019*).

*Tidal Forces and the Roche Limit*

Stellar tidal forces must overcome the planetesimal's self-gravity to facilitate disruption into dusty debris; such tidal forces are also important in preventing the resulting debris disk from reaccreting into coherent objects (*von Hippel et al. 2007; van Lieshout et al. 2018).* This condition defines the Roche Limit ($d_R$) and derivations of its value for strengthless rigid objects

$$d_R = R_{star}\sqrt[3]{\frac{2\rho_{star}}{\rho_{object}}} \tag{1}$$

where $R_{star}$ and $\rho_{star}$ are the radius and density of the host star, and $\rho_{object}$ is the bulk density. Given the high density of white dwarfs, this equation can be readily rewritten to require only the mass of the white dwarf ($M_{star}$) and density of orbiting material

$$d_R = \sqrt[3]{\frac{3M_{star}}{2\pi\rho_{object}}}. \tag{2}$$

Note that these equations assume negligible centripetal acceleration on the planetesimal's surface (e.g., a non-rotating body).

White dwarfs are known to have masses ranging from 0.17 $M_\odot$ (*Kilic et al. 2007*) to ~1.3 $M_\odot$, although the distribution is sharply peaked between 0.5 - 0.7 $M_\odot$[1] (*McCleery et al. 2020*). If we assume that the material comprising disrupted planetesimals is typical asteroidal material (densities of ~2000 - 3500 kg/m$^3$), we can estimate the rigid body Roche limit for most white dwarfs with debris disks to be between 0.0034 - 0.0046 AU (or equivalently, 0.74 - 1.00 $R_\odot$).

[1] DB white dwarfs tend to be more massive than DA white dwarfs, however are numerically dominated by the DA white dwarfs and their mass distribution (*e.g., McCleery et al. 2020*).

Another possible choice is the fluid Roche limit for objects subject to fluid-like deformation. This may be a reasonable model for such planetesimals, as comparable asteroidal objects within our own Solar System are subject to significant tidal deformation (*Walsh & Richardson, 2006; 2008; Zhang & Michel, 2020*). Nevertheless, we choose to employ the rigid body Roche limit, which is a more stringent limit that lies nearer to the star, resulting in a more conservative model for the behavior of white dwarf dusty debris disks.

*Material Vaporization and the Sublimation Radius*

The inner edge of a dusty debris disk must lie sufficiently far from its host star for a given material/composition to stay cool enough to remain solid. Inside of this material-specific "vaporization radius" (or "sublimation radius"), solid materials would rapidly enter the gas phase, dissipating the debris disk. Thermodynamically, illuminated materials predominantly cool via thermal radiation if they are sufficiently cold. However, as materials heat up (i.e., approach the host star), they begin to cool predominantly through overcoming the heat of sublimation (sublimative cooling). The location of the sublimation radius, and thus the inner physical limit of a dusty debris disk, is the location at which materials become sufficiently hot to transition from radiative to sublimative cooling.

We determine the location of this sublimation radius using the material sublimation model of *Steckloff et al. (2015)*, which solves the energy balance equation at the illuminated surface of an object, to determine equilibrium temperature, and thus the heat flux resulting from each mechanism. To compute stellar luminosity, we assume blackbody emission from a spherical white dwarf, whose radius we compute from its mass using the model of *Carvalho et al. (2015)*. This sublimation model relies on five parameters of a species to describe its sublimative behavior (latent heat of sublimation, molar mass, sublimation coefficient, and a reference pressure/temperature pair), which can be determined from laboratory studies. This model already includes the olivine species fayalite and forsterite. To consider the behavior of other common species, we add metallic iron and the pyroxene mineral enstatite to this sublimation model. Since enstatite breaks down into forsterite and silica at high temperatures (*Tachibana et al. 2002*), this modeled behavior of enstatite is merely an estimate of its sublimative behavior. See Appendix A for details of this sublimation model, white dwarf radius model, and the inclusion of iron and enstatite into this model.

Using this model, we can numerically solve for the equilibrium temperature as a function of white dwarf luminosity (i.e., temperature) and astrocentric distance (see Figure 1). We then use this surface temperature to compute the amount of energy lost through radiative and sublimative heat loss, and thus the ratio between them. We define the sublimation radius as the location where radiative and sublimative heat loss are equal, which occurs at temperatures of 2560 K, 2100 K, 2160 K, and 1540 K, for forsterite, fayalite, iron, and enstatite, respectively. Inside of the sublimation radius dust will efficiently sublimate away. Significant sublimation can nevertheless occur outside of this definition of the sublimation radius; if one were to instead define the sublimation radius as the point where sublimative heat loss is a half or full order-of-magnitude lower than the radiative heat loss, the sublimation radii would be, respectively, only ~10% or ~25% larger. In any case, the sublimation radius represents the innermost possible edge of a dusty debris disk.

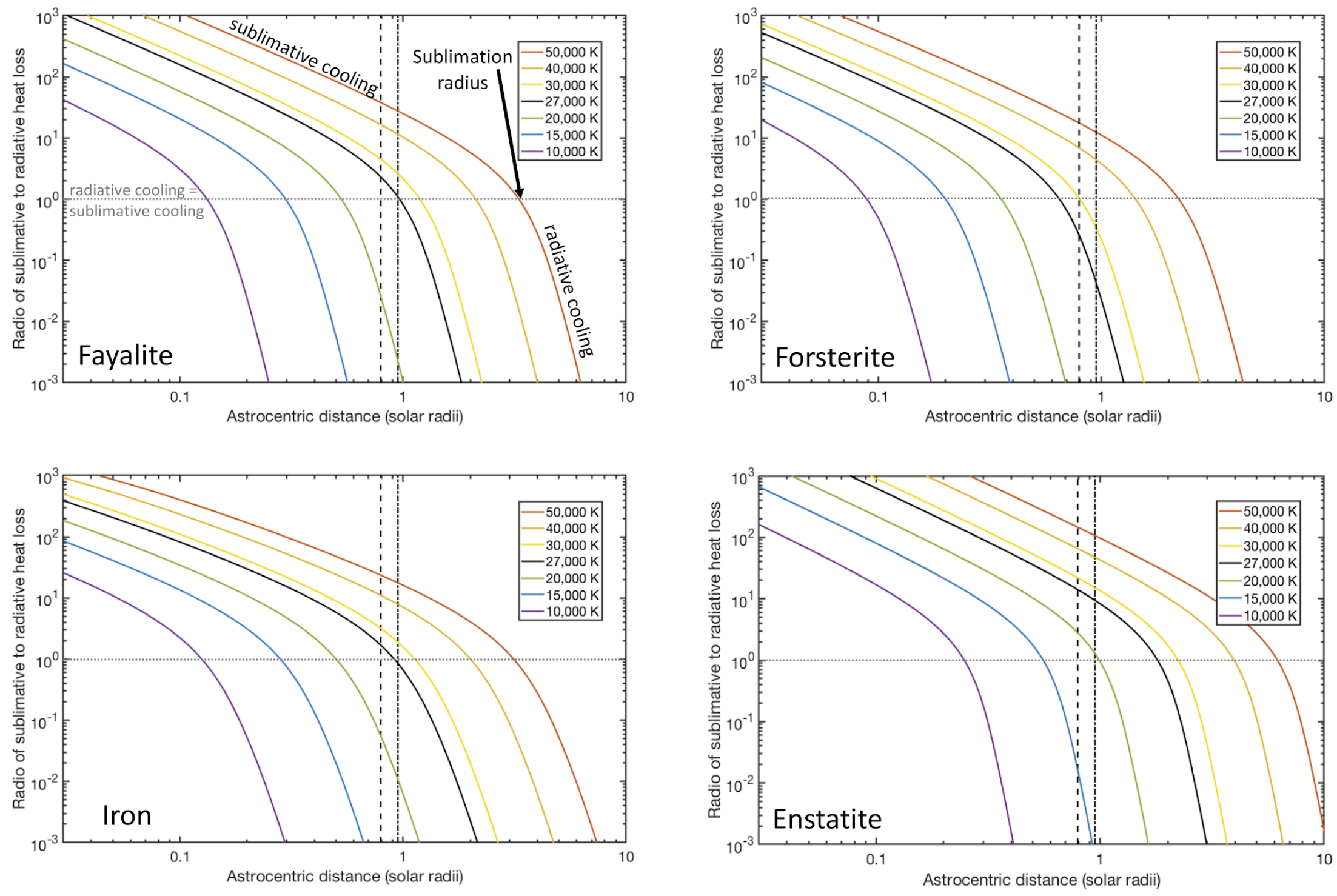

**Figure 1:** *Scaling relation (ratio of sublimative to radiative heat loss) for the olivine species fayalite and forsterite, metallic iron, and the pyroxene species enstatite as a function of distance and temperature of 0.6 solar mass white dwarf.* The region of sublimative and radiative cooling regimes are denoted on the fayalite plot, which applies to all four species. Dashed and dash-dotted lines denote the rigid body Roche limit for densities of 3500 and 2000 kg/m$^3$ respectively. The transition point from radiative to sublimative cooling when this ratio approaches 1 (horizontal dotted grey line) creates an inflection point that represents the sublimation radius, and thus the inner edge of any disk of dusty debris composed of that species. These four species are major components of chondritic and achondritic materials, and represent likely major components of a dusty debris disk.

## III. White Dwarf Cooling and the Onset of Debris Disk Stability

To remain stable against either subliming to gas[2] or reccreting into a coherent object (*van Lieshout et al. 2018*), dusty debris typically needs to both reside outside of the sublimation radius, yet interior to the Roche limit (or cross the Roche limit on eccentric orbits; *Malamud and*

[2] This model neglects other processes that can produce vapor, such as hypervelocity impacts or sputtering

*Perets, 2020*). This requires that the sublimation radius of the materials comprising a disrupted planetesimal lie interior to the Roche limit. Unlike the Roche limit of the dusty debris, which is independent of the temperature/luminosity of the white dwarf (and therefore remains stationary), the sublimation radius strongly depends on the temperature/luminosity of the white dwarf, and therefore recedes starward as the white dwarf cools. Young, hot white dwarfs have high luminosities, resulting in sublimation radii that lie far outside the Roche limit. Thus, were an inwardly scattered planetesimal to cross inside of the Roche limit and disrupt, the resulting materials would rapidly sublimate to gas, destroying the grains that would otherwise form a dusty debris disk. Conversely, old, cool white dwarfs have low luminosities, resulting in sublimation radii that lie well interior to the Roche limit, allowing for a region of space in which an inwardly scattered planetesimal can disrupt and form a debris disk without experiencing sublimative disruption. Thus, as white dwarfs cool, the sublimation radius migrates closer to the star, and will eventually recede inside of the Roche limit, forming a region of thermal stability for dusty debris disks (*von Hippel et al. 2007; Koester et al. 2014*).

Therefore a critical white dwarf effective temperature exists for which the sublimation radius equals the Roche limit; this temperature represents the warmest possible white dwarf temperature for which dusty debris disks can form (see figure 2). This critical temperature is different for different materials, due to their differing volatilities. Additionally, a white dwarf's radius (and therefore blackbody luminosity) and Roche limit depend on mass; thus this critical temperature also depends on the mass of the white dwarf.

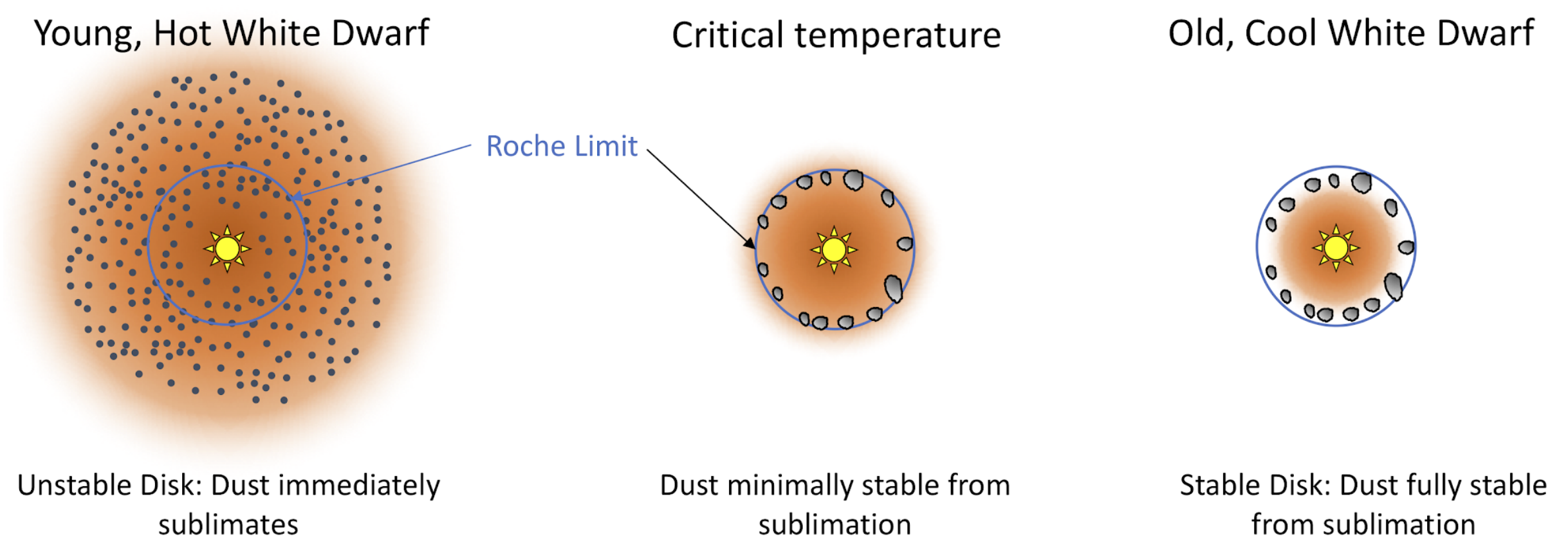

**Figure 2:** *Cartoon of the onset of dusty debris disk stability*. (*Left)* The sublimation radius (edge of red sublimation region) exceeds the Roche limit (blue) around young, hot white dwarfs, preventing the formation of dusty debris disks that are stable against sublimative destruction. *(Middle)* As the star cools, the sublimation radius recedes inward, and eventually crosses the Roche limit; at this critical temperature, dust in a debris disk would be minimally stable against sublimation. *(Right)* As the star continues to cool, the stable region for debris disks expands to include regions closer to the star.

We use our model to compute the sublimation radius for common planetary materials (forsterite, fayalite, enstatite, and iron) as a function of white dwarf temperature and mass, and compare this with the Roche limit (see figure 3). We find that, for a typical white dwarf (mass of 0.6 $M_{solar}$), the critical temperature, and thus the onset of dusty debris disk formation, lies between ~25,000K and 32,000K, depending on planetesimal density and composition[3]. These critical temperatures change little for the majority of white dwarfs, which have masses between ~0.5 - 0.7 $M_{solar}$. This is remarkably consistent with the observed onset of dusty debris disks at ~27,000K (*e.g., Xu & Jura, 2012; Xu et al. 2015; Li et al. 2017; Debes et al. 2019*), suggesting that indeed the sublimative behavior of the planetary materials controls the onset of dusty debris disk formation. This finding provides further support for the two-step dusty debris disk formation model of white dwarfs, which was used to derive the conditions for the critical temperature. Thus, although the inward-scattering flux of planetesimals around young-hot white dwarfs may be inherently higher due to the presence of recently destabilized planetary systems, the intense thermal environment inside of the Roche limit nevertheless prevents the formation of dusty debris disks.

This model also rules out formation of water ice-rich dusty debris disks from comet-like or KBO-like objects. The onset of stability for such water-ice rich disks around a 0.6 $M_{\odot}$ white dwarf results in a critical temperature below ~2,700 K (assuming object density of 1000 kg/m$^3$); this results in a white dwarf cooling age longer than the age of the Universe (*Richer et al. 2000*). Nevertheless, the detected signatures of accreted water in some white dwarf photospheres (*Farihi et al. 2013; Raddi et al. 2015; Hoskin et al. 2020*) could result from dusty debris disks containing hydrated minerals rather than free water (*e.g., Hoskin et al. 2020*).

Our model suggests that there is a strong stellar mass dependence for the critical temperature. More massive white dwarfs have smaller radii (*Carvalho et al. 2015*), and are therefore less luminous for a given temperature, leading to smaller sublimation radii. At the same time, more massive white dwarfs have larger Roche limits. These two features suggest that more massive white dwarfs have higher critical temperatures. For example a 1.3 $M_{\odot}$ white dwarf (which is below the Chandrasekhar limit of 1.4 $M_{\odot}$, where the white dwarf collapses into a neutron star) has a Roche limit between 1.16 and 1.40 $R_{\odot}$ (for planetesimal densities between, respectively, 3500 and 2000 kg/m$^3$). This corresponds to critical temperatures of 73,000K, 60,000K, 62,000K, and 44,000K for forsterite, fayalite, iron, and enstatite respectively, for a planetesimal with a density of 3500 kg/m$^3$. For a planetesimal with a density of 2000 kg/m$^3$, these critical temperatures are 81,000K, 66,000K, 68,000K, and 48,000K for forsterite, fayalite, iron, and enstatite respectively. Disks of extremely refractory materials such as corundum could exist around even hotter white dwarfs, although the relative rarity of such materials makes such disks unlikely to be found. However, while dusty debris disks around massive, hot white dwarfs are possible, they are unlikely: large white dwarfs are rare, they cool quickly, and their high luminosities would make the infrared excess signature of the dusty debris disk difficult to detect with current instruments.

[3] This temperature range is significantly warmer than previous calculations, which found a critical temperature of 15,000 - 22,000 K (*von Hippel et al. 2007; Koester et al. 2014*); this discrepancy is due to the use of a simplified thermal stability model (*von Hippel et al. 2007*) that assumes a sublimation temperature, rather than solve for it.

Conversely, previous efforts have suggested that warm white dwarfs are likely to vaporize dusty materials, and would instead host gaseous debris disks (*Koester et al. 2014; Manser et al. 2020*). Our model suggests that this is only part of the story, as the sublimative destruction of dusty debris depends on stellar luminosity, which is a function of white dwarf effective temperature and radius/mass. The larger radii of less massive white dwarfs leads to higher luminosities for a given temperature. As a result, dusty materials will sublimate more readily around less massive white dwarfs. Such thermal environments may even be extreme enough to destroy dusty materials far enough from the white dwarf to prevent accretion, consistent with the PG 0010+280 system (*Xu et al. 2015*). Nevertheless, it is presently unclear whether observations of white dwarf gaseous debris disks (*e.g., Melis et al. 2020; Dennihy et al. 2020; Steele et al. 2020; Manser et al. 2020*) are consistent with this expectation.

Lastly, no dusty debris disks have been observed around massive white dwarfs ($M_\star$>0.9 $M_\odot$; *Kilic et al. 2008*). This could be due to small number statistics, as such massive white dwarfs are quite rare (*e.g., McCleery et al. 2020; Hollands et al. 2020*). Alternatively, the absence of observed dust is consistent with the two-step model of debris-disk formation, which requires Poynting-Robertson drag to circularize the orbits of dusty debris from disrupted planetesimals. The Poynting-Robertson drag force ($F_{P-R}$) depends linearly on stellar luminosity ($L_\star$) and orbital speed (*Burns et al. 1979*), which itself scales as the square of stellar mass ($M_\star$), thus

$$F_{P-R} \sim L_\star \sqrt{M_\star}. \qquad (3)$$

The resulting Poynting-Robertson drag force in a 0.9 $M_\odot$ system is only ~1/3 as strong as that of an otherwise equivalent 0.6 $M_\odot$ white dwarf; with relative strength decreasing with increasing mass. Furthermore, this force must remove more specific orbital angular momentum from the debris in massive systems to circularize their orbits. This inherently slower dynamical evolution suggests that dusty debris disks around more massive white dwarfs form much more slowly, with the resulting debris migrating inward toward the star much more slowly. This makes dusty debris and dusty debris disks around more massive white dwarfs inherently harder to both form and be detected, which may provide a natural explanation for the observation that no white dwarfs more massive than ~0.9 $M_\odot$ are known to host dusty debris disks.

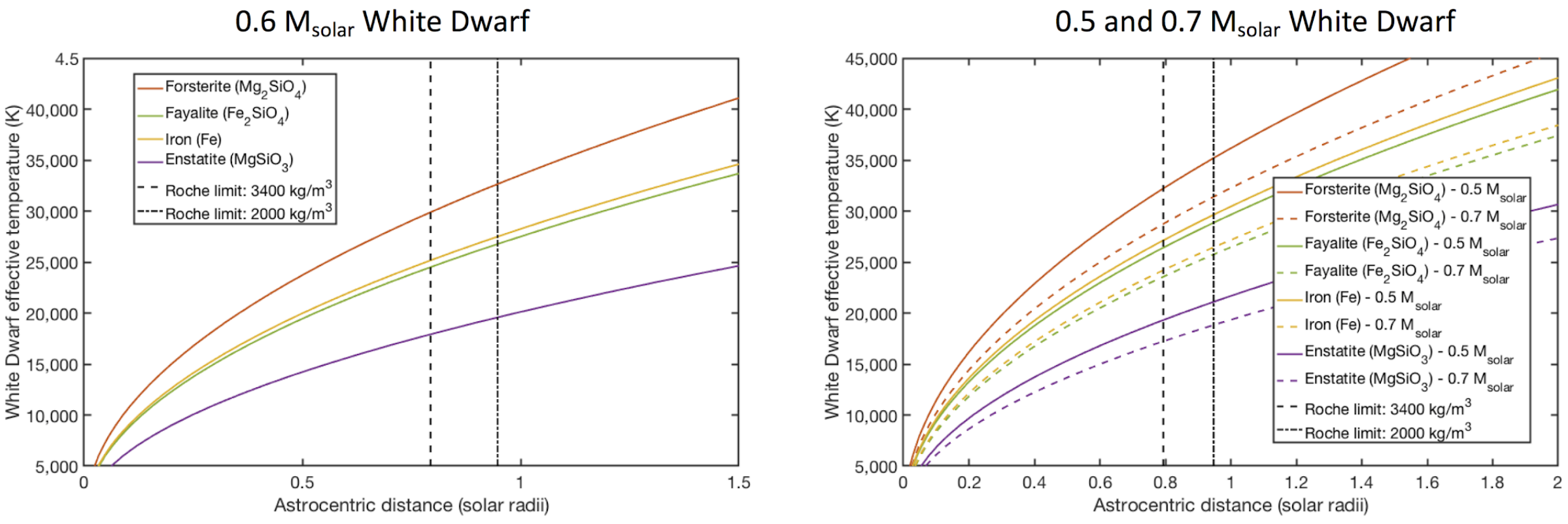

**Figure 3:** Relationship between white dwarf temperature and sublimation radius for forsterite, fayalite, iron, and enstatite. We compute the radius of transition from radiative to sublimative cooling (the sublimation radius), and thus the inner edge of a dusty debris disk, as a function of white dwarf temperature for these four species for the most probable white dwarf mass (0.6 $M_{solar}$), and the most probable range of white dwarf masses (0.5 and 0.7 $M_{solar}$). Dashed and dash-dotted lines denote the rigid body Roche limit for densities of 3500 and 2000 kg/m$^3$ respectively.

## IV. Consistency with Observed White Dwarf Debris Disks

Because white dwarf dusty debris disks must lie between the inner sublimation radius and outer Roche limit to be stable, we can compare these model results with observation-based estimates of white dwarf debris disks. We use our model to calculate the inner and outer edges (sublimation radii and rigid Roche limits) of the region of stability for dusty debris disks in the G29-38, WD 2115-560, GD 362, GD 56, PG1541+651, LSPM J0207+3331, and PG 0010+280 systems (see Table 1); LSPM J0207+3331 and PG 0010+280 are particularly compelling as they represent, respectively, the coolest and warmest white dwarfs known to host dusty debris disks. We model the observed infrared excesses of these systems as a flat disk, which is not a unique solution due to degeneracies between disk width and inclination and limited photometric quality. Nevertheless, we generally find good agreement between these flat disk modeled observations and our stability model.

The inner edge of the disk in the G29-38 system, which analysis of observations suggests lies between 0.15 and 0.28$R_\odot$(*Zuckerman & Becklin, 1987; Jura, 2003; Reach et al. 2005*), agrees closely with the inner sublimation radius of both iron (0.15 $R_\odot$) and the silicate mineral olivine (fayalite: 0.16 $R_\odot$; forsterite: 0.11 $R_\odot$). This is consistent with the composition of the dusty debris, which shows strong olivine (*Reach et al. 2005; von Hippel et al. 2007*) and iron-rich pyroxene signatures (*Reach et al. 2009*), suggesting that a planetesimal with a composition typical of Solar System asteroids disrupted to form this disk.

Similarly, observations suggest that the disk in the GD 362 system is silicate-rich (*Zuckerman et al. 2007*) and lies between 0.08 and 0.50 $R_\odot$(*von Hippel et al. 2007*). This is also in excellent agreement with the inner sublimation radius of iron (0.04 $R_\odot$) and olivine (fayalite: 0.04 $R_\odot$, forsterite: 0.03$R_\odot$), which again suggests that a planetesimal of typical composition disrupted to form the observed dusty debris disk. The inner edge of the dusty debris disk in the WD 2115-560 system is likewise consistent with iron and olivine.

The disks in the GD 56 and PG1541+651 systems have observed inner edges that lie interior to sublimation radii of iron and fayalite. This suggests that, if the planetesimals that disrupted to form these disks were typical olivine-rich asteroids, the olivine would likely be dominated by the forsterite end-member, which is more refractory than fayalite. Thus, the forsterite sublimation radius (0.18 - 0.19 $R_\odot$ and 0.10 $R_\odot$, respectively) lies interior to the inner edge of the disks (0.21 $R_\odot$ and 0.13 $R_\odot$, respectively). Although iron-dominated asteroids and meteoroids are common in our solar system, the iron sublimation radius lies exterior to the inner edge of the debris disks, ruling out an iron-dominated planetesimal.

The LSPM J0207+3331 (6120 K, 3 Gyr cooling age) system is unique; it is the coldest, oldest white dwarf known to host a dusty debris disk, and this disk appears to have a sizable gap within it (*Debes et al. 2019*). The outer edge of the disk in the LSPM J0207+3331 system lies near the system's Roche limit; the inner edge of the disk lies at the sublimation radius for both fayalite and iron, and outside that of forsterite. While this does not exclude forsterite as a component of the disk, the compelling agreement of the inner disk edge with the sublimation radii of fayalite and iron suggest that the inner disk is composed of these materials. The apparent gap in this disk is curious and unexpected, and, analogous to protoplanetary disks, may point to the presence of a dense planet clearing a gap along its orbit from within the disk (as proposed for the SDSS J122859.93+104032.9 system; *Manser et al. 2019*) or a planet sitting outside the Roche limit opening a gap via resonant dynamics. Similar dynamical processes may be at work in dusty debris disks around white dwarfs and the ~3 Gyr cooling age of this white dwarf provides ample time for such dynamical processes to occur.

Alternatively, it is also possible that the LSPM J0207+3331 system is actually hosting two debris disks that formed from the disruption of two independent planetesimals. Dusty debris disks are thought to dissipate through radiative and thermal processes such as Poynting-Roberson drag and sublimation; leading to estimates of dust disk lifetimes of order ~$10^6$ yrs (*Rafikov, 2011; Girven et al. 2012; Metzger et al. 2012*). Because stellar luminosities, and thus radiative and thermal processes, are so low around such a cool, old white dwarf, disk lifetimes in such systems can be more than an order of magnitude longer than systems around warmer ($T_{eff}$ > ~10,000 K) white dwarfs (*Rafikov, 2011*), potentially long enough to preserve a remnant disk from a previous disruption event. If these longer lifetimes are comparable to (or longer than) the interval between typical debris disk forming events, then such gaps or multiple-disk systems may be relatively common in old white dwarf systems. Future observations may be able to test this hypothesis.

On the other temperature extreme, the inner disk radius of the PG 0010+280 system, although poorly constrained to 0.13 - 0.65 $R_{\odot}$(*Xu et al. 2015*), agrees with forsterite's sublimation radius of 0.66 $R_{\odot}$. However, observations were unable to detect spectral signatures of white dwarf pollution (*Xu et al. 2015*) to confirm this composition; this may be the first known white dwarf with a dusty debris disk that lacks detectable spectral pollution. Alternatively, the observed infrared excess could be the result of a 1300 K blackbody, such as a substellar companion or gas giant, rather than a dusty debris disk (*Xu et al. 2015*). Assuming the observed infrared excess is the result of a dusty debris disk, the extreme thermal environment combined with the relatively large distance that such pollution would have to traverse to accrete onto the white dwarf may preclude such pollution. In this case, it is likely that other young, hot white dwarfs may not show spectral pollution in spite of hosting dusty debris disks.

G29-38, GD 56, GD 362, PG 1541+651, and LSPM J0207+3331 all have disks with inner edges near the sublimation radii of various species. One interpretation is that these disks are evolutionarily mature compared to the disk in the WD 2115-560 system, and have had sufficient time for Poynting-Robertson drag or viscous spreading to bring the detectable inner edges of these disks to the sublimation radius. The GD 362 system appears to have a debris disk with an inner edge near the sublimation radius of forsterite, suggesting a forsterite-rich disk. Nevertheless, a detailed discussion of the relative timescales of these processes is beyond the scope of this manuscript.

| **System** | **Observed Properties** | | | | **Disk Inner Edge/ Sublimation Radius ($R_{\odot}$)** | | | | **Rigid Roche Limit ($R_{\odot}$)** | |
|---|---|---|---|---|---|---|---|---|---|---|
| | **$T_{eff}$ (K)** | **Mass ($M_{\odot}$)** | **Cooling Ages (MYR)** | **Modeled Disk range ($R_{\odot}$)** | Forsterite | Fayalite | Iron | Enstatite | $\rho_{object}$ = 2000 kg/m$^3$ | $\rho_{object}$ = 3500 kg/m$^3$ |
| G29-38 | 11357[a] | 0.62 | 450 | 0.15 - 0.28[b,c,d,e] | 0.11 - 0.13 | 0.16 - 0.19 | 0.15 - 0.18 | 0.30 - 0.35 | 0.92 - 0.99 | 0.77 - 0.82 |
| WD | 9674[f] | 0.59[b] | 640 | 0.17 - 0.32[b] | 0.08 | 0.11 | 0.11 | 0.21 | 0.98 | 0.81 |
| 2115-560 | | | | | | | | | | |
| GD 362 | 9740[g] | 0.71[l] | | 0.08 - 0.50[b] | 0.072 | 0.11 | 0.10 | 0.20 | 1.00 | 0.83 |
| GD 56 | 151510[b] | 0.62[b] | 200 | 0.21 - 0.58[b] | 0.18 - 0.19 | 0.27 - 0.29 | 0.26 - 0.27 | 0.51 - 0.54 | 0.90 - 0.93 | 0.75 - 0.77 |
| PG 1541+651 | 11,278[h] | 0.60[h] | 440 | 0.13 - 0.36[i] | 0.10 | 0.16 | 0.15 | 0.29 | 1.01 | 0.84 |
| LSPM J0207+3331 | 6120[j] | 0.69[j] | 3000 | 0.047 - 0.21, ~0.94 | 0.030 | 0.044 | 0.041 | 0.081 | 0.99 | 0.82 |
| PG 0010+280 | 24206[h] | 0.52[h] | 17 | Inner:0.13 - 0.65[k] Outer:0.52 - 1.3[k] | 0.66 | 0.98 | 0.93 | 1.8 | 0.93 | 0.77 |

**Table 1:** *Table of observed and calculated properties of select systems.* We list the properties of the observed white dwarf systems hosting dusty debris disks, along with our computed sublimation radii within those systems for various planetary materials and rigid-body Roche limits for typical endmember planetesimal densities. The observed locations of the disks are in alignment with the calculated sublimation radii (inner disk edge) and Roche radii (outer disk edge). Refs: [a] Kepler & Nelan, 1993 [b] von Hippel et al. 2007 [c] Zuckerman & Becklin, 1997 [d] Jura, 2003 [e] Reach et al. 2005 [f] Koester et al. 2005 [g] Gianninas et al. 2004 [h] Gianninas et al. 2011 [i] Kilic et al. 2012 [j] Debes et al. 2019 [k] Xu et al. 2015 [l] Kilic et al. 2008

## V. Conclusions

We introduce a sophisticated material sublimation model to the study of white dwarf dusty debris disks. With this model, we find that the interplay between a white dwarf's material-dependent sublimation radius (which recedes as the white dwarf cools) and the static Roche limit provide a natural explanation for the observed restriction of dusty debris disks to white dwarfs that have cooled to below ~27,000 K (*e.g., Xu & Jura, 2012; Xu et al. 2015; Li et al. 2017; Debes et al. 2019*). This agreement provides further support for the two-step model of white dwarf dusty debris disk formation (*Graham et al. 1990; Jura, 2003; Debes et al. 2012; Veras et al. 2014; Malamud and Perets, 2020*). This also rules out the possibility of pure water ice-rich debris disks (not to be confused with disks rich in hydrated minerals). We apply our model to a selection of observed white dwarf systems, and find good agreement between the observed disk location and composition and our model expectations (Table 1). Our model suggests that the observed locations of the dusty debris disks within the PG 0010+280, GD 56, and PG 1541+651 systems may indicate that they were formed from a planetesimal composed of forsterite-dominated olivine, a silicate mineral. Finally, we find that cooler white dwarfs such as LSPM J0207+3331 are capable of, and perhaps likely to, host multiple dusty debris disks from disruptions of different inwardly scattered planetesimals, due to the inherently long histories and slow radiatively driven dynamical evolutions of these systems.

## VI. Acknowledgements

We wish to thank the anonymous referee, whose comments greatly improved the clarity, structure, and context of this work. This work was supported in part by NASA award 80NSSC20K0267 (J.K. Steckloff, A. Steele, E.R. Adams, and S.A. Jacobson) and NASA contract NNM10AA11C issued through the New Frontiers Program (A. Springmann).

## VII. Appendix A: Material Sublimation Model

We use the sublimation model of *Steckloff et al. (2015*) to identify the location of the sublimation radius for the species forsterite, fayalite, iron, and enstatite. This model solves the energy balance equation at the illuminated surface of an object, to determine equilibrium temperature, and thus the heat flux resulting from each mechanism is

$$I_{stellar} = I_{radiative} + I_{sublimative} \tag{A1}$$

$$I_{stellar} = \frac{L_{stellar}}{4\pi r_a^2} \tag{A2}$$

$$I_{radiative} = 2\epsilon\sigma T^4 \tag{A3}$$

$$I_{sublimative} = \lambda_{sub} \overline{m}_{(T)} \tag{A4}$$

where $I_{stellar}$ is the stellar insolation flux, $I_{radiative}$ is the radiative heat flux, $I_{sublimative}$ is the sublimative heat flux, $L_{stellar}$ is the stellar luminosity, $r_a$ is the astrocentric distance, $\epsilon$ is the blackbody emissivity of the surace, $\sigma$ is the Stefan-Boltzmann constant, $T$ is the surface temperature, $\lambda_{sub}$ is the latent heat of sublimation of the surface material, and $\overline{m}$ is the sublimative mass flux from the surface. *Steckloff et al. (2015*) used the Knudsen-Langmuir equation (*Langmuir, 1913)* and the Clausius-Clapyron phase change relation to derive an expression for sublimative mass flux

$$\overline{m} = \alpha_{(T)} \sqrt{\frac{m_{mol}}{2\pi R T}} P_{(T)} \tag{A5}$$

$$P_{(T)} = P_{ref} e^{\frac{\lambda_{sub}}{R}\left(\frac{1}{T_{ref}} - \frac{1}{T}\right)} \tag{A6}$$

where $\alpha_{(T)}$ is the temperature-dependent sublimation coefficient (typically of order 1; *Langmuir, 1913*), $m_{mol}$ is the molar mass of the sublimating species, $R$ is the ideal gas constant, $T$ is the surface temperature, and $P_{ref}$ and $T_{ref}$ are a laboratory-determined vapor pressure and temperature pair, which anchors the phase-change relationship (*Steckloff et al. 2015*). Thus, the sublimative behavior of any chemical species is determined by five quantities: sublimation coefficient ($\alpha_{(T)}$), molar mass ($m_{mol}$), latent heat of sublimation ($\lambda_{sub}$), and a reference pressure and temperature ($P_{ref}$and $T_{ref}$). Typically, the sublimation coefficient is assumed to have a value of 1, however we use experimental data where available; we use the temperature-dependent sublimation coefficient for water from *Gundlach et al. (2011)*, and we derive the temperature dependence for the sublimation coefficient of Forsterite from published values (see Appendix B).

This model was originally developed to study sublimation of comet ices ($H_2O$, $CO_2$, and CO), and was later expanded to include the olivine species forsterite and fayalite (*Steckloff et al. 2015; Steckloff & Jacobson 2016*). To study the sublimative behavior of dusty debris disks, we expand this model to include other common chemical species found in asteroidal materials (e.g., metallic iron and pyroxene). To include metallic iron into our sublimation model, we fit the experimental vapor pressure and temperature data from *Desai (1986)* to equation A6 to obtain a best fit. Although the best fit to the heat of sublimation for iron is 415.47 kJ/mol at 298.15 K, this heat of sublimation drops at temperature increases (*Desai, 1986*), we find that the best fit to the *sublimation* behavior of iron is to use a heat of sublimation of 340.8 kJ/mol, and a pressure and temperature reference of 14.52 Pa and 1867 K respectively. We also assume a sublimation coefficient of 1 for all temperatures.

We could not identify comparable laboratory data of pyroxenes (e.g., enstatite) sufficient for a comparable fitting of vapor-solid equilibrium. One major complication is the thermal breakdown of pyroxenes such as enstatite ($MgSiO_3$), which loses silicon dioxide ($SiO_2$) and forms a layer of forsterite ($Mg_2SiO_4$) on the subliming surface (*Tachibana et al. 2002).* This results in a complicated diffusive process, which generally breaks the simple sublimating surface approximation that we have adopted in our model. Nevertheless, we include an approximation to the behavior of enstatite using the activation energy that *Tachibana et al.* (*2002)* identified for this process of 457 kJ/mol, and a vapor pressure - temperature pair of 100 Pa at 1420 K from *Lewis (1973*). We again assume a sublimation coefficient of 1 for all temperatures.

We compare this model with the sublimation model of Rafikov and Garmilla (*2012*), and find that they produce similar temperature dependencies of sublimative mass flux as a function of temperature. For iron, we found that the models agree very closely when the $T_0$ parameter in the Rafikov and Garmilla (2012) model is changed to 40,989 K, to match the equivalent $\lambda_{sub}/R$ parameter in the Steckloff et al. model. Greater inconsistencies were present for olivine, largely due to the Steckloff et al. model considering the different olivine species forsterite and fayalite, while the Rafikov and Garmilla (*2012*) model only considered a generic olivine (not broken down by olivine species). We also attempted to compare with the sublimation model of Shestakova et al. *(2019)*, but we were unable to reconstruct their model.

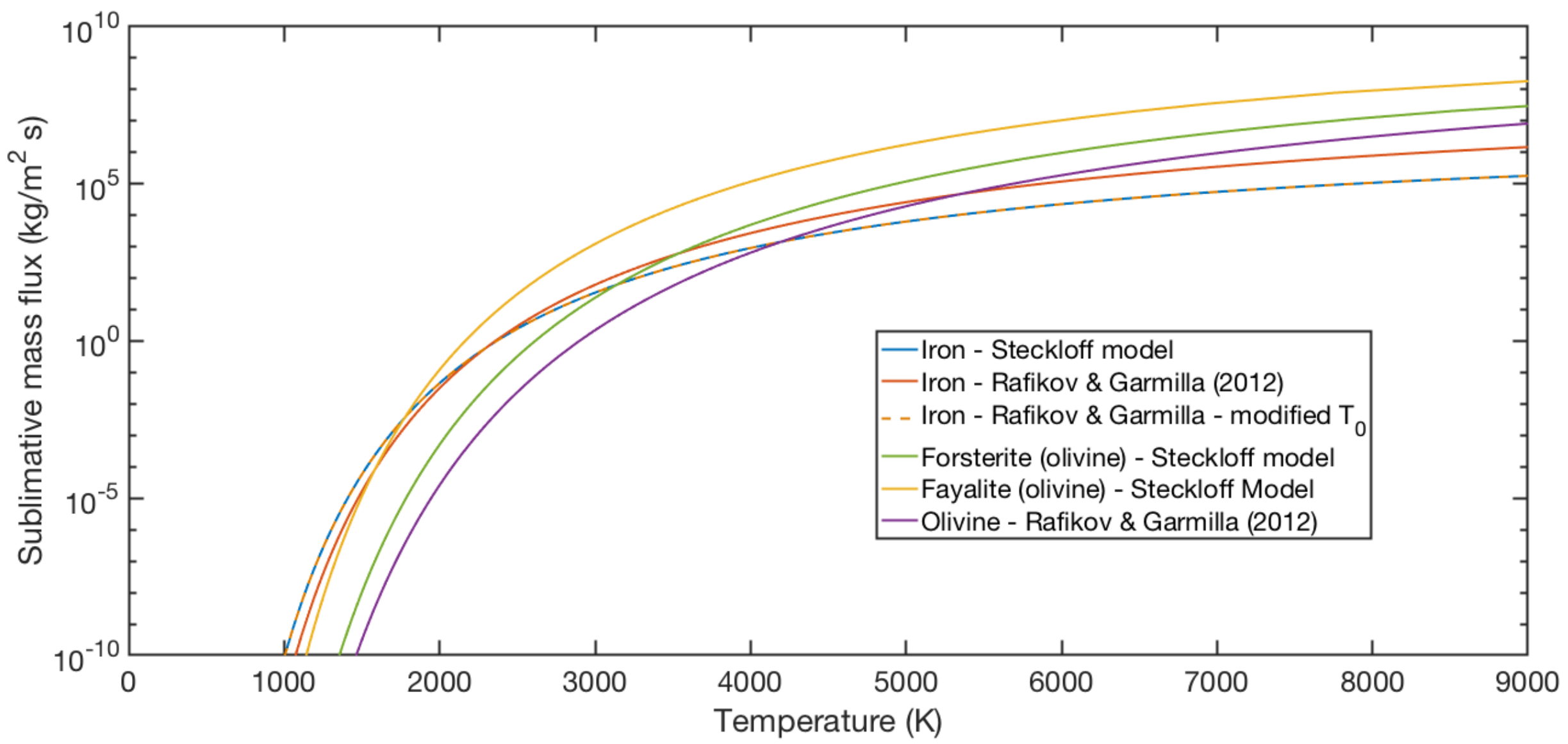

**Figure A1:** *Comparison to the model of Rafikov & Garmilla (2012)*. Comparing the temperature dependence of sublimative mass flux in these models for both iron and olivine (or olivine species forsterite and fayalite). These models agree to within two orders of magnitude. For iron, these two models agree closely when the $T_0$ parameter in the Rafikov and Gamilla (2012) model is changed to match the Steckloff et al. model.

We calculate the luminosity of a white dwarf ($L_{stellar}$) in equation 2 by computing the blackbody radiative flux for a specified temperature, and then integrating over the surface area of the star. We assume the star is a sphere, thus the surface area depends only on the stellar radius ($R_{star}$), which we approximate from the stellar mass ($M_{star}$)

$$\frac{M_{star}}{M_{\odot}} = \frac{aR_{star} + b}{e^{cR^2_{star}} + d} \quad \text{(A7)}$$

where $a$=2.325 x 10$^{-5}$ km$^{-1}$, $b$= 0.4617, $c$ = 7.277 x 10$^{-9}$ km$^{-2}$, $d$ = -0.644 (*Carvalho et al. 2015*). We numerically solve this equation to solve for stellar radius as a function of mass, and thus stellar luminosity.

# VIII. Appendix B: Fitting The Sublimation Coefficients for Forsterite

Forsterite sublimation has been extensively studied experimentally, to determine the temperature dependence of its sublimation coefficient (see all cited experiments). However, experimental temperatures rates do not exceed 2073 K, which is less than the maximum temperatures that we simulate here. We therefore extrapolate the temperature dependence of this behavior using the curve fitting toolbox in MATLAB.

| Temperature (K) | Sublimation coefficient | Reference |
|---|---|---|
| 1673 | 0.04 | Hashimoto, 1990 |
| 1773 | 0.038 | Wang et al. 1999 |
| 1873 | 0.053 | Wang et al. 1999 |
| 1953 | 0.074 | Wang et al. 1999 |
| 1973 | 0.12 | Hashimoto, 1990 |
| 2023 | 0.097 | Wang et al. 1999 |
| 2073 | 0.12 | Wang et al. 1999 |

**Table B1:** *Experimental sublimation coefficients of Forsterite*. We searched the literature for experimentally derived sublimation coefficients of forsterite in a vacuum. This table lists published figures from the literature.

From experiments with $H_2O$ (Gundlach et al. 2011), we expect the temperature dependence of forsterite's sublimation coefficient ($\alpha_{(T)}$) to follow the functional dependence

$$\alpha_{(T)} = \frac{a}{1+e^{b(\frac{1}{T}-c)}} + d \quad \text{(A8)}$$

where a, b, c, and d are experimentally determined coefficients. In the literature, we compiled experimentally derived sublimation coefficients for forsterite sublimation into a vacuum (see table A1), and fitted them to determine the coefficients a-d. Our best fit curve produced the coefficients listed in Table A2 (see figure A2).

| coefficient | value |
|---|---|
| a | 0.1569 |
| b | 44570 K |
| c | $4.906 \times 10^{-4}$ $K^{-1}$ |
| d | 0.03679 |

**Table B2:** *Best fit coefficients for temperature dependence of forsterite sublimation coefficient*

We find that the resulting curve (Figure A1) has a low temperature asymptote (≲1600 K) of 0.37 and a high temperature asymptote (≳2500 K) of 0.19, with rapid change between these two regions. This curve broadly agrees with other references (Hashimoto, 1990; Kuroda & Hashimoto, 2002; Nagahara & Ozawa, 1996; Takigawa et al. 2009; Tsuchiyama et al. 1999; Tsuchiyama et al. 1998) that show the sublimation coefficient of forsterite is between ~0.01–0.1, providing confidence that our extrapolation is reasonably accurate. We include this result into our sublimation dynamics for forsterite (see Appendix A).

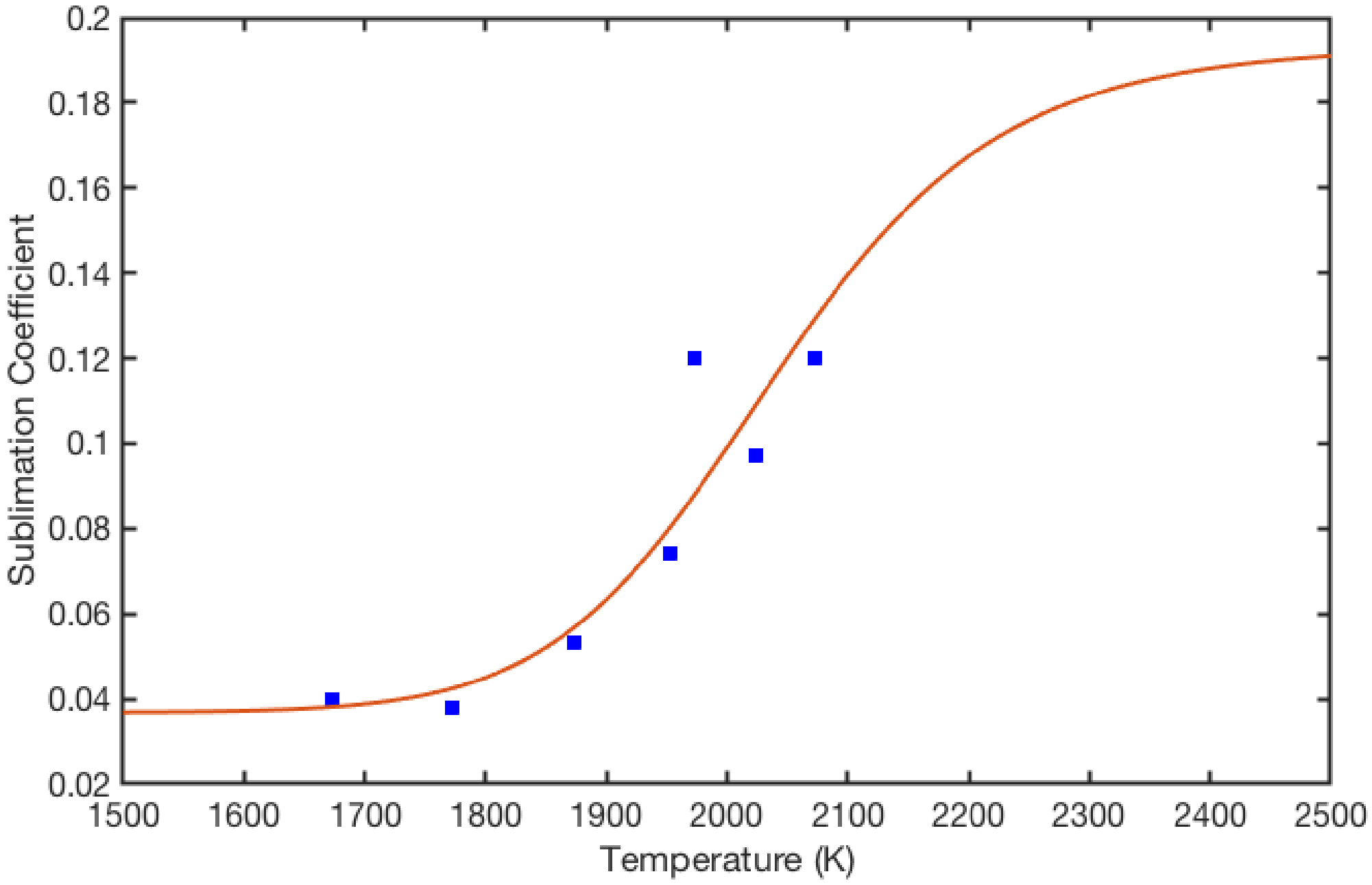

**Figure B2:** *Fitted extrapolation of forsterite sublimation coefficient.* We fitted the experimental data, to obtain this empirical temperature dependence on the sublimation coefficient of forsterite.

## IX. References

Barstow, M.A., Barstow, J.K., Casewell, S.L., et al. 2014, *MNRAS,* **440**, 1607

Bonsor. A., Mustill, A.J., & Wyatt, M.C. 2011, *MNRAS,* **414**, 930

Bonsor, A., & Wyatt, M.C. 2012, MNRAS, **420**, 2990

Burns, J.A., Lamy, P.L., & Soter, S. 1979, *Icarus,* **40**, 1

Canup, R.M. 2010, *Nature,* 468, 943

Carvalho, G.A., Marinho Jr, R.M., & Malheiro, M. 2015, *J. of Physics: Conference Series* **630**, 012058

Debes, J.H. & Sigurdsson, S. 2002, *ApJ,* **572**, 556

Debes, J.H., Hoard, D.W., Kilic, M. et al. 2011, *ApJ*, **729**, 4

Debes, J.H., Walsh, K.J., Stark, C. 2012, *ApJ,* **747**, 148

Debes, J.H., Thévenot, M., Kuchner, M.J., et al. 2019, *ApJ,* **872**, L25

Dennihy, E., Xu, S., Lai, S. *et al.* 2020, *ApJ,* **905**, 5

Desai, P.D. 1986, *J. Phys. Chem. Ref. Data,* **15**, 967

Farihi, J., Becklin, E.E., & Zuckerman, B. 2008, *ApJ,* **681**, 1470

Farihi, J. 2016, *New Astron. Rev.,* **71**, 9

Farihi, J., Gänsicke, B.T., Koester, D. 2013, *Science*, **342**, 218

Frewen, S.F.N., & Hansen, B.M.S. 2014, *MNRAS,* **439**, 2442

Gianninas, A., Dufour, P., & Bergeron, P. 2004, ApJ, 617, L57

Gianninas, A., Bergeron, P., & Ruiz, M.T. 2011, *ApJ,* **743**, 138

Girven, J., Brinkworth, C.S., Farihi, J. et al. 2012, *ApJ*, **749**, 154

Graham, J.R., Mattherw, K., Neugebauer, G., et al. 1990, *ApJ,* **357**, 216

Gundlach, B., Skorov, Y.V., & Blum, J. 2011, *Icarus,* **213**, 710

Hansen, B.M.S., Kulkarni, S., & Wiktorowicz, S. 2006, *Astron. J.,* **131**, 1106

Hashimoto, A. 1990, *Nature,* **347**, 53

von Hippel, T., Kuchner, M.J., & Kilic,M. 2007, *ApJ,* **662**, 544

Hollands, M. A., Tremblay, P.-E., Gaensicke, B.T., et al. 2020, *Nature Astronomy*, 4, 663

Hoskin, M.J., Toloza, O., Gänsicke, B.T., et al. 2020, *MNRAS*, **499**, 171

Jura, M. 2003, *ApJ,* **584**, L91

Kepler, S.O., & Nelan, E.P. 1993, *AJ,* **105**, 608

Kilic, M., Prieto, C.A., Brown, W.R., et al. 2007, *ApJ,* **660**, 1451

Kilic, M., Patterson, A.J., Barber, S., et al. 2012, *MNRAS,* **419**, L59

Kilic, M., Thorstensen, J.R., & Koester, D. 2008, *ApJ,* **689**, L45

Koester, D., Rollenhagen, K., Napiwotzki, R., et al. 2005, *A&A,* **432**, 1025

Koester, D., Gänsicke, B.T., & Farihi, J. 2014, *A&A,* **566**, A34

Kuroda, D., & Hashimoto, A. 2002, *Antarct. Meteorite Res.,* **15**, 152

Langmuir, I. 1913, *Phys. Rev.* 2, 329

Lewis, J.S. 1973, *Annu. Rev. Phys. Chem,* **24**, 339

Li, L., Zhang, F., Kong, X., et al. 2017, *ApJ,* **836**, 71

Van Lieshout, R., Kral, Q., Charnoz, S., et al. 2018, *MNRAS,* **480**, 2784

Malamud, U., Perets, H.B., 2020 *MNRAS*, **492**, 5561

Malamud, U., Grishin, E., Brouwers, M. 2021, *MNRAS*, **501**, 3806

Manser, C.J., Gänsicke, B.T., Eggl, S. et al. 2019 *Science,* **364**, 66

Manser, C.J., Gänsicke, B.T., Fusillo, G. et al. 2020 *MNRAS*, **493**, 2127

McCleery, J., Tremblay, P.-E., Gentile, F., et al. 2020 *MNRAS,* **499**, 1890

Melis, C., Klein, B., Doyle, A.E., et al. 2020, *ApJ,* **905**, 56

Metzger, B.D., Rafikov, R.R., Bochkarev, K.V., *MNRAS,* **423**, 505

Mustill, A.J., Villaver, E., Veras, D., et al. 2018, *MNRAS,* **476**, 3939

Nagahara, H., & Ozawa, K. 1996, *Geochimica et Cosmochimica Acta,* **60**, 1445

Nordhaus, J.& Spiegel, D. S. 2013, *MNRAS,* **432**, 500

Raddi, R., Gänsicke, B.T., Koester, D., et al. 2015, *MNRAS*, **450**, 2083

Rafikov, R.R., 2011, *ApJL*, **732**, L3

Rafikov, R.R. & Garmilla, J. A. 2012, *ApJ*, **760**, 123

Reach, W.T., Kuchner, M.J., Hippel, T.v., et al. 2005, *ApJ,* **635**, L161

Reach, W.T., Lisse, C., von Hippel, T. et al. 2009, *ApJ*, **693**, 697

Richer, H.B., Hansen, B., Limongi, M., et al. 2000, *ApJ,* **529**, 318

Rocchetto, M., Farihi, J., Gänsicke, B.T., et al. 2015, *MNRAS,* **449**, 574

Shestakova, L.I., Demchenko, B.I., Serebryanskiy, A.V. 2019, *MNRAS*, **487**, 3935

Springmann, A., Lauretta, D.S., Klaue, B., et al. 2019, *Icarus,* **324**, 104

Steckloff, J.K., Johnson, B.C., Bowling, T., et al. 2015, *Icarus,* **258**, 430

Steckloff, J.K. & Jacobson, S.A. 2016, *Icarus,* **264**, 160

Steele, A., Debes, J., Xu, S., et al. 2020, *ApJ.* Accepted for Publication

Tachibana, S., Tsuchimyama, A., & Nagahara, H. 2002, *Geochimica et Cosmochimica Acta,* **66**, 713

Takigawa, A., Tachibana, S., Nagahara, H., et al. 2009, *ApJ,* **707**, L97

Tsuchiyama, A., Tachibana, S., & Takahashi, T. 1999, *Geochimica et Cosmochimica Acta* **63** 2451

Tsuchiyama, A., Tachibana, S., & Takahashi, T. 1998, *Mineralogical Journal* **20**, 113

Villaver, E. & Livio, M. 2007, *ApJ,* **661**, 1192

Veras, D., Mustill, A.J., Bonsor, A. et al. 2013, *MNRAS,* **431**, 1686

Veras, D., Leinhardt, Z. M., Bonsor, A., et al. 2014, *MNRAS,* **445**, 2244

Veras, D., Leinhardt, Z. M., Eggl, S., et al. 2015, *MNRAS,* **451**, 3453

Walsh, K. J. & Richardson, D.C. 2006, *Icarus* **180**, 201

Walsh, K. J. & Richardson, D. C. 2008, *Icarus,* ***193***, 553

Wang, J., Davis, A.M., Clayton, R.N., et al. 1999, *Geochimica et Cosmochimica Acta* **63**, 953

Wilson, T.G., Farihi, J., Gänsicke, B.T., Swan, A. 2019, *MNRAS*, **487**, 133

Xu, S. & Jura, M. 2012, *ApJ,* **745**, 88

Xu, S., Jura, M., Pantoja, B., et al. 2015, *ApJ L* **806**, L5

Zuckerman, B., & Becklin, E.E. 1987, *Nature,* **330**, 138

Zuckerman, B., Koester, D., Melis, C., et al. 2007, *ApJ,* **671**, 872

Zuckerman, B., Koester, D., Reid, I. N., et al. 2003, *ApJ*, **596**, 477

Zuckerman, B., Melis, C., Klein, B., et al. 2010, *ApJ*, **722**, 725

Zhang, Y. & Michel, P. 2020, *A&A,* **540**, A102